\documentclass[reprint,aps,prb,superscriptaddress,twocolumn,longbibliography,floatfix]{revtex4-2}
\input{glyphtounicode}
\usepackage[T1]{fontenc}
\usepackage[utf8]{inputenc}
\usepackage[utf8]{inputenc}
\DeclareUnicodeCharacter{0328}{\c{}}
\usepackage{graphics,epsfig,amsfonts,amssymb,amsmath,color}
\usepackage{accents}
\usepackage{bbm}
\usepackage{color}
\usepackage{comment}
\usepackage{enumitem}
\usepackage{natbib}
\usepackage[colorlinks=true,linkcolor=blue,citecolor=blue,urlcolor=blue,hypertexnames=false]{hyperref}
\usepackage{textcomp}
\usepackage{setspace}

\newcommand{\ba}{\begin{array}{rl}}
	\newcommand{\ea}{\end{array}}

\newcommand{\change}[1]{\textcolor{black}{{#1}}}

\begin{document}
% \linenumbers

%\author{}
%		%\email[]{Your e-mail address}
%		%\homepage[]{Your web page}
%		%\thanks{}
%		%\altaffiliation{}
%		\affiliation{}

%%%%% from here

\title{A 2$\times$2 quantum dot array in silicon with fully tunable pairwise interdot coupling}

\author {Wee Han Lim}
\email{wee.lim@unsw.edu.au}
\affiliation {School of Electrical Engineering and Telecommunications, University of New South Wales, NSW 2052, Sydney, NSW, Australia.}
\affiliation {Diraq, Sydney, NSW, Australia.}
\author {Tuomo Tanttu}
\affiliation {School of Electrical Engineering and Telecommunications, University of New South Wales, NSW 2052, Sydney, NSW, Australia.}
\affiliation {Diraq, Sydney, NSW, Australia.}
\author {Tony Youn}
\affiliation {School of Electrical Engineering and Telecommunications, University of New South Wales, NSW 2052, Sydney, NSW, Australia.}
\author {Jonathan Yue Huang}
\affiliation {School of Electrical Engineering and Telecommunications, University of New South Wales, NSW 2052, Sydney, NSW, Australia.}
\author {Santiago Serrano}
\affiliation {School of Electrical Engineering and Telecommunications, University of New South Wales, NSW 2052, Sydney, NSW, Australia.}
\affiliation {Diraq, Sydney, NSW, Australia.}
\author {Alexandra Dickie}
\affiliation {School of Electrical Engineering and Telecommunications, University of New South Wales, NSW 2052, Sydney, NSW, Australia.}
\affiliation {Diraq, Sydney, NSW, Australia.}
\author {Steve Yianni}
\affiliation {School of Electrical Engineering and Telecommunications, University of New South Wales, NSW 2052, Sydney, NSW, Australia.}
\affiliation {Diraq, Sydney, NSW, Australia.}
\author {Fay E. Hudson}
\affiliation {School of Electrical Engineering and Telecommunications, University of New South Wales, NSW 2052, Sydney, NSW, Australia.}
\affiliation {Diraq, Sydney, NSW, Australia.}
\author {Christopher C. Escott}
\affiliation {School of Electrical Engineering and Telecommunications, University of New South Wales, NSW 2052, Sydney, NSW, Australia.}
\affiliation {Diraq, Sydney, NSW, Australia.}
\author {Chih Hwan Yang}
\affiliation {School of Electrical Engineering and Telecommunications, University of New South Wales, NSW 2052, Sydney, NSW, Australia.}
\affiliation {Diraq, Sydney, NSW, Australia.}
\author {Arne Laucht}
\affiliation {School of Electrical Engineering and Telecommunications, University of New South Wales, NSW 2052, Sydney, NSW, Australia.}
\affiliation {Diraq, Sydney, NSW, Australia.}
\author {Andre Saraiva}
\affiliation {School of Electrical Engineering and Telecommunications, University of New South Wales, NSW 2052, Sydney, NSW, Australia.}
\affiliation {Diraq, Sydney, NSW, Australia.}
\author {Kok Wai Chan}
\affiliation {School of Electrical Engineering and Telecommunications, University of New South Wales, NSW 2052, Sydney, NSW, Australia.}
\affiliation {Diraq, Sydney, NSW, Australia.}
\author {Jesús D. Cifuentes}
\affiliation {School of Electrical Engineering and Telecommunications, University of New South Wales, NSW 2052, Sydney, NSW, Australia.}
\affiliation {Diraq, Sydney, NSW, Australia.}
\author{Andrew S. Dzurak} 
\email{a.dzurak@unsw.edu.au}
\affiliation {School of Electrical Engineering and Telecommunications, University of New South Wales, NSW 2052, Sydney, NSW, Australia.}
\affiliation {Diraq, Sydney, NSW, Australia.}

\date{\today}

\begin{abstract}

\change{Recent advances in semiconductor spin qubits have enabled linear arrays with more than ten qubits. Scaling to two-dimensional (2D) arrays is essential for fault-tolerant implementations but introduces significant fabrication challenges due to the increased density of gate electrodes. Moreover, implementing two-qubit entanglement control requires the addition of interstitial exchange gates between quantum dots in this dense gate structure. In this work, we present a 2D array of silicon metal-oxide-semiconductor (MOS) quantum dots with tunable interdot coupling between all neighboring dots. Characterized at 4.2 K, the device exhibits exceptional tunability, supports the formation and isolation of both double-dot and triple-dot configurations, and achieves tunnel coupling control spanning up to 30 decades per volt. These results provide critical technical feedback and a foundational benchmark for advancing MOS spin qubit technology into the 2D regime. }

\end{abstract}

%Recent advances in semiconductor spin qubits have achieved linear arrays exceeding ten qubits. Moving to two-dimensional (2D) qubit arrays is critical for fault-tolerant implementations but poses substantial fabrication challenges. Enabling control of nearest-neighbor entanglement requires interstitial exchange gates between quantum dots. This work presents a 2D array of silicon metal-oxide-semiconductor (MOS) quantum dots with tunable interdot coupling between all adjacent dots. Characterized at 4.2 K, the device demonstrates the formation and isolation of double-dot and triple-dot configurations. We show control of all nearest-neighbor tunnel couplings spanning up to 30 decades per volt through the interstitial exchange gates.

% a pivotal step towards the fabrication of spin qubit arrays with MOS technolog
% Manually show keywords

\maketitle

\section{Introduction}

%\noindent
Recent research efforts have already demonstrated significant progress in semiconductor-based spin qubits, showcasing high-fidelity operations \cite{madzik2022precision,xue2022quantum,noiri2022fast,mills2022two,Weinstein2023universal,tanttu2024assessment}, high temperature operations \cite{Petit2018,Yang2020,Petit2020,camenzind2022hole,huang2024high}, as well as increasing qubit counts from devices made in both academic\cite{philips2022universal} and foundry\cite{george202412} cleanrooms. One key advantage of a silicon-based platform is the ability to leverage the semiconductor advanced manufacturing capabilities to scale up the number of qubits to millions for a utility-scale quantum computer. 

Most of these results have, however, been reported in linear qubit arrays with the current record lying at 12 spin qubits measured in a linear 12-dot device~\cite{george202412}. Advancement from 1D to 2D arrays of quantum dots is necessary for the development of universal quantum computing architectures compatible with error correction methods such as surface code~\cite{fowler_surface_2012}. Recent proposals indicate that a 2$\times$N quantum dot array is enough for a first implementation of error correction \cite{siegel2024towards}, considering that one of the rows is integrated by physical qubits while the second row enables entanglement via spin-shuttling.

Two-dimensional arrays of semiconductor quantum dots have been implemented using heterostructures in GaAs~\cite{mukhopadhyay20182, Mortemousque_GaAs3x3_2021}, Si/SiGe~\cite{unseld20232d, wang2024highly}, and Ge/SiGe~\cite{wang2024operating, borsoi2024shared, hendrickx2021four, zhang2024universal}. \change{A key feature in the most advanced implementations is the ability to control tunnel coupling, which is essential for optimal spin readout and enabling entanglement of nearest neighbors via exchange interactions. Modern devices incorporate interstitial exchange gates between quantum dots for this purpose. High-fidelity two-qubit gates were achieved shortly after integrating these gates into the devices~\cite{xue_quantum_2022, mills_two-qubit_2022} and their implementation in 2D arrays have led to the first demonstrations of 2D spin qubit quantum processors~\cite{hendrickx2021four}.}

%the most successful ones, the implementation of two qubit
%gates followed soon after the observation of exchange in-
%teractions6–8, with confirmed realizations of high fidelity
%$two qubit gates (>99%) in spin qubits in silico

\change{High-fidelity spin qubits have also been demonstrated in MOS quantum dots at the Si/SiO$_2$ interface~\cite{huang2024high,tanttu2024assessment}. These devices are fabricated with materials and processes that are standard in the semiconductor industry~\cite{stuyck2024cmos}, which opens up substantial advantages for scalability, including the possibility of fabricating spin qubit arrays at semiconductor foundries~\cite{steinacker2024300mmfoundrysilicon} and co-integrating them with conventional electronics on the same chip~\cite{Vandersypen2017,Veldhorst2017}.}

%\change{Only few of these demonstrations have led to complete 2D spin qubit systems, also involving spin-readout, control and pairwise entanglement}Modern devices also require the inclusion of interstitial exchange gates between dots to control the interdot tunnel coupling, which is necessary for optimal spin readout and to enable entanglement of nearest neighbors via exchange interactions~\cite{tanttu2024assessment}.

\change{However, a key limitation of MOS spin qubit technology has been its smaller quantum dot size. The pitch of a state-of-the-art MOS quantum dot device is less than half that of equivalent devices in Si/SiGe or Ge/SiGe~\cite{burkard2023semiconductor}, demanding very fine lithography with a low error margin.  The inclusion of interstitial exchange gates further reduces the already narrow gate pitch by half, pushing the limits of academic cleanrooms and semiconductor foundries~\cite{stuyck2024cmos}. Although there have been demonstrations of 2D quantum dot arrays in silicon MOS~\cite{gilbert2020single,duan2020remote,ansaloni2023gate}, none have yet demonstrated tunnel coupling control between all quantum dot pairs.}

Here, we report the fabrication and measurement of a 2$\times2$ quantum dot array in a silicon MOS architecture with tunable tunnel coupling between all quantum dot pairs (FIG. \ref{Fig1}(a)). \change{The device is fabricated with electron beam lithography (EBL) using up to four metalization layers to integrate interstitial exchange gates}.  We characterize quantum dot formation at 4.2 K and demonstrate double dot and triple dot configurations in the device. We also characterize the tunnel rate control of exchange gates growth in different metalization layers. Based on these results and further modeling and simulation, we provide valuable insights into the development of 2D spin qubit architectures.

% . array of gates.  which is necessary for optimal Pauli-spin readout and for controlling nearest-neighbor entanglement via exchange interactions~\cite{tanttu2024assessment}. 
% Modern devices use interstitial exchange gates for this purpose \cite{hendrickx2021four} . 

% However, the smaller gate-pitches impose and additional challenge to  cleanrooms and in semiconductor foundries.}

\section{Device Design and Fabrication}\label{Device design and fabrication}

\begin{figure*}%[ht!]
\centering
\includegraphics[width=6.1in]{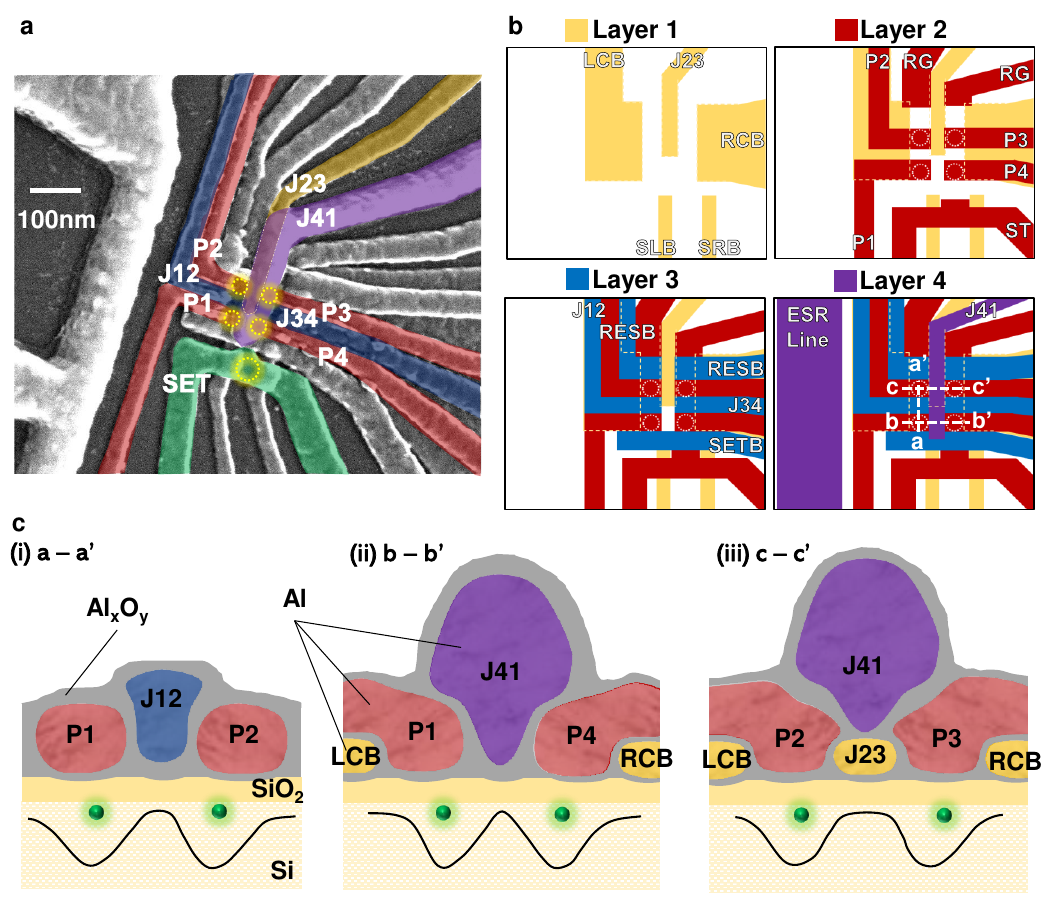}
\caption{Device architecture overview. (a) False-coloured SEM image of an identical device used in the experiment. Four quantum dots are formed under respective P1$-$P4 gates with four individual J-gates to control the interdot coupling between pairs of dots. A single-electron transistor (SET) is fabricated nearby to detect the charge occupancy in the dots. (b) Schematic representation of the multi-level gate stack device, illustrating the layering of gates in order. (c) Schematic cross-section of the device showing the position of a barrier gate (J12, J41 and J23) fabricated in between two plunger gates, as indicated by the white dashed lines in (b). The geometry of the barrier gates has a strong influence on the conduction band profiles.}
\label{Fig1}
\end{figure*}

Figure~\ref{Fig1}(a) is a scanning electron micrograph (SEM) of a device nominally identical to the one measured in this work. The device consists of a four-layer aluminum gate stack~\cite{Angus2007,Lim2009-2}, fabricated using EBL, thermal deposition of Al and lift-off, on top of thermally-grown high-quality SiO$_2$ on a $^{\mathrm{nat}}$Si substrate. Aluminum oxide, Al$_x$O$_y$ is \change{thermally} formed at each layer to electrically isolate the subsequent layer of Al gates. Figure~\ref{Fig1}(b) shows the schematic of laying different gates at each layer forming a gate-stack of 2$\times$2 quantum dots with a single-electron transistor (SET). Starting from layer 1 (yellow), we define quantum dot confinement gates (LCB and RCB), SET barrier gates (SLB and SRB), and a barrier gate between dot 2 and dot 3 (J23). Layer 2 (red) consists of plunger gates (P1, P2, P3 and P4), reservoir gate (RG) and an SET top gate (ST). Layer 3 (blue) comprises barrier gates (J12 and J34), reservoir barrier gates (RESB) and an SET barrier gate (SETB). The uppermost layer (purple) consists of a barrier gate (J41) and an on-chip microwave line (ESR). The ESR line is not used in this work. The Al metal thickness on Layers 1, 2, 3 and 4 are 16, 28, 29 and 100 nm, respectively.

In this design, we define all four plunger gates in the same layer but J-gates are defined in different layers. This allows us to investigate the influence of J-gate on interdot tunnel coupling based upon its geometry. Figure~\ref{Fig1}(c) illustrates the cross-sectional schematics of the possible J-gate geometries. From previous transmission electron microscopy (TEM) analyses of multiple device geometries~\cite{lim2009electrostatically,tanttu2024assessment,cifuentes2024bounds}, we see the following trends in gate layer stacking. As shown in FIG.~\ref{Fig1}(c)(i), J12 (also J34) is fabricated in the $\sim$20 nm gap between the adjacent plunger gates and hence forming a slightly rounded profile on the gate oxide. In comparison, J41 in FIG.~\ref{Fig1}(c)(ii) is formed with a slightly pointed profile on top of the gate oxide. On the other hand, J23 is fabricated in layer 1, making it a relatively larger barrier gate ($\sim$30 nm). The geometry and position of the J-gates will have a strong influence on the conduction band profiles, as indicated in FIG.~\ref{Fig1}(c).

The RG gates extend over a phosphorus doped region (not shown here) to allow the accumulation of electron layers when a positive voltage, above threshold, is applied. This forms the source of electrons to be loaded into the quantum dots underneath the P1, P2, P3 and P4 gates to form a 2$\times$2 quantum dot array. Both RESB gates act as tunnel barriers for loading of electrons from the reservoirs to the P2 and P3 dots. The J-gates are used to control the interdot tunnel coupling between the four quantum dots. For instance, J12 controls the coupling between P1 and P2 dots while J23 controls the coupling between P2 and P3 dots.

\begin{figure*}[ht!]
\centering
\includegraphics[width=6.1in]{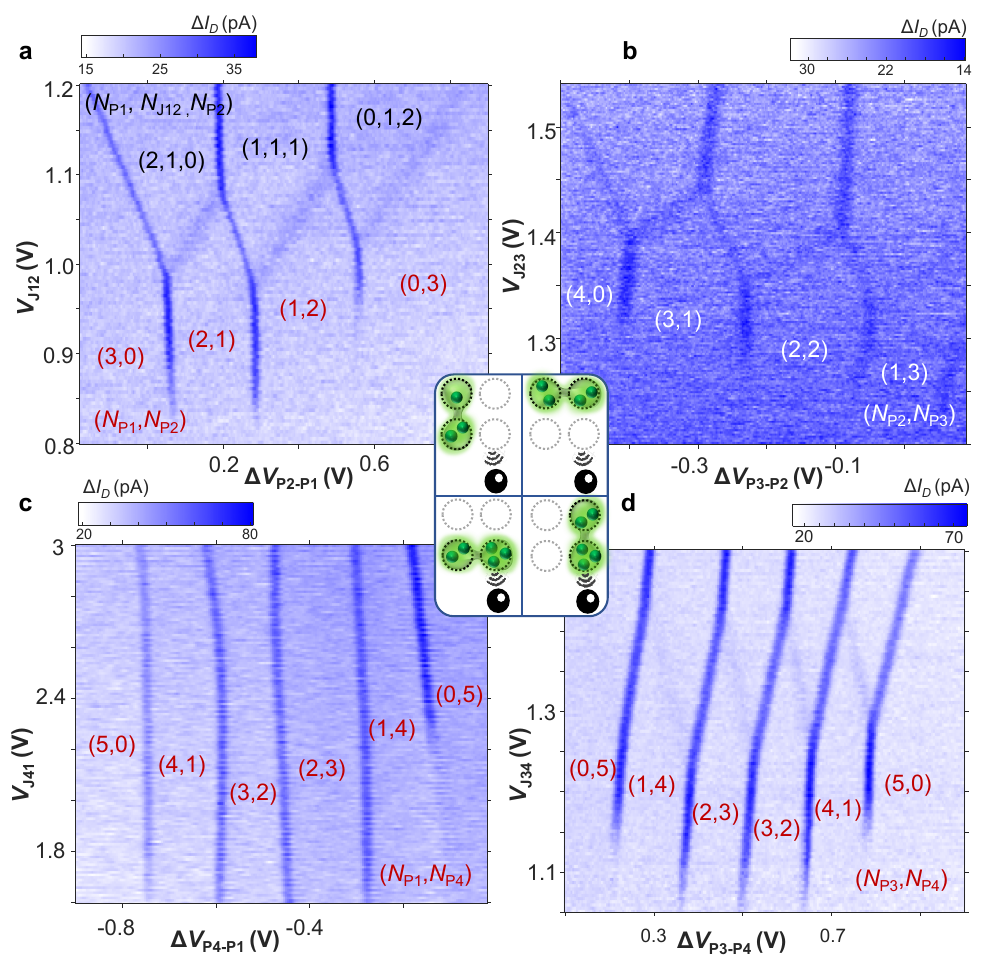}
\caption{Isolated-mode charge stability diagrams for a 2$\times$2 quantum dot array system showing the electron occupancy on each pairwise dots. The measured differential current, $\Delta$$I_\mathrm{D}$ is obtained from the SET charge sensor as a function of J-gate voltage, $V_\mathrm{J}$ and plunger gate detuning voltage, $\Delta$$V_\mathrm{P}$. Each line in the stability diagram indicates a charge transition when an electron tunneling event occurs between the pairwise dots. (a) Charge stability diagram for the P1$-$P2 pairwise dots showing the electron occupancy on dot P1 and P2 in ($N_\mathrm{P1}$,$N_\mathrm{P2}$). Three electrons were loaded into the double dots formed under gates P1 and P2. At low $V_\mathrm{J12}$ voltage (<1V), a double dot is formed and the electrons are loaded from P1 to P2 dots, one by one as the detuning voltage $\Delta$$V_\mathrm{P2-P1}$ is increased.  At $V_\mathrm{J12}$ > 1V, gate J12 forms a dot underneath and a triple quantum dot charge configuration is measured, with their electron occupancy indicated in ($N_\mathrm{P1}$,$N_\mathrm{J12}$,$N_\mathrm{P2}$). Similarly, (b), (c) and (d) show the charge stability diagram for P2$-$P3, P1$-$P4 and P3$-$P4 pairwise dots, respectively.}
\label{Fig2}
\end{figure*}

\section{Double Dot Charge Characteristics}\label{Double dot characteristics}

\begin{figure*}[ht!]
\includegraphics[width=6.365in]{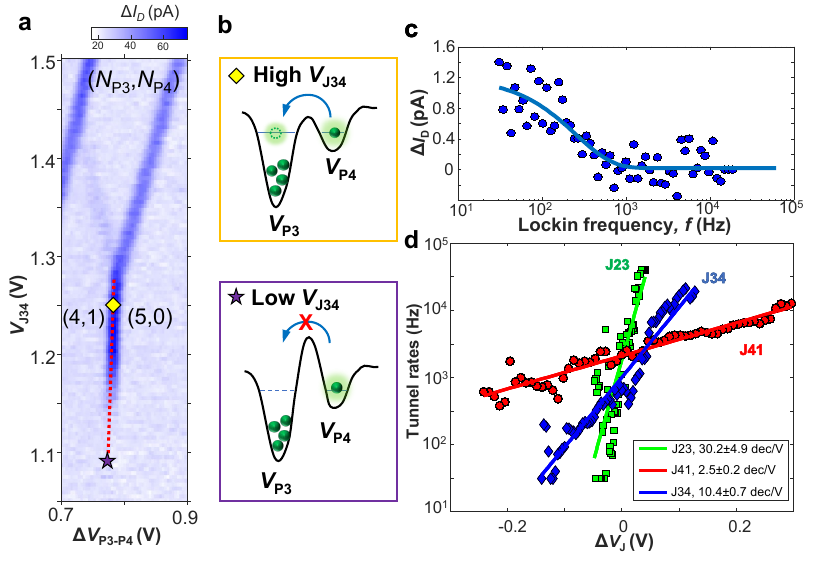}
%\vspace{0.15in}
\caption{Tunnel rates measurement between pairwise dots as a function of J-gate voltage, $V_\mathrm{J}$.
(a) Focused region of FIG.~\ref{Fig2}(d) showing the electron transition of P3$-$P4 dots in the (4,1)--(5,0) electron configuration. (b) Schematic showing the conduction band profiles for P3$-$P4 dots at high and low J34 voltages. (c) Measurement of the SET differential current, $\Delta$$I_\mathrm{D}$ as a function of the gate-pulsed lock-in frequency, $f$ from 31 Hz$-$61 kHz to probe the electron tunnel rates between P3$-$P4 dots along the transition line, as indicated by the red dashed line. (d) Tunnel rates as a function of $\Delta$$V_\mathrm{J}$. The rate of increase of tunnel rates is shown in the legend.}
\label{Fig3}
\end{figure*}

In this section, we present the electrical measurement of the 2D array of quantum dots sensed via an SET charge sensor at 4.2 K. We use a gate-pulsed lock-in charge sensing technique~\cite{Yang2012} to characterize the charge transitions and electron occupancy of each quantum dot in isolated-mode operation~\cite{eenink2019tunable,Yang2020}. In isolated mode, the RESB and RG gate voltages are reduced to below threshold after the dots are loaded with the desired number of electrons in order to pinch off the tunnel rates to the electron reservoir.

Figure~\ref{Fig2} shows the charge stability diagrams of the 2$\times$2 quantum dot array for (a) P1$-$P2, (b) P2$-$P3, (c) P1$-$P4 and (d) P3$-$P4 pairwise dots. The lines in the stability diagram are charge transitions when electron tunneling events occur between the pairwise dots. Between the transition lines, the number of electrons in each P dot is fixed and stipulated as $N_\mathrm{P}$. In Figure ~\ref{Fig2}(a), three electrons are loaded into the double dots formed under gates P1 and P2. Their electron occupancy is indicated in ($N_\mathrm{P1}$,$N_\mathrm{P2}$). At low $V_\mathrm{J12}$ (< 1V), a double dot is formed and the electrons are shuttled from P1 to P2 dots, one by one, as the detuning voltage $\Delta$$V_\mathrm{P2-P1}$ is increased. As $V_\mathrm{J12}$ is further reduced, the barrier between P1 and P2 dots increases resulting in lower tunnel rates and hence the sensing signal $\Delta$$I_\mathrm{D}$ diminishes. Conversely, when we further increase $V_\mathrm{J12}$ (> 1V in this case), instead of creating a tunnel barrier, an unintended dot is created under the J12 gate. At this point, we form a triple quantum dot system ($N_\mathrm{P1}$,$N_\mathrm{J12}$,$N_\mathrm{P2}$) where the tunnel barriers in between P1, J12 and P2 are caused by the thin layer of Al$_x$O$_y$. 

It is worth noting that the formation of J-dots is also observed under J23 (FIG.~\ref{Fig2}(b)) and J34 (FIG.~\ref{Fig2}(d)), but not J41 (FIG.~\ref{Fig2}(c)). This is owing to the geometry and position of J41 being narrower and defined higher up in the stack compared to other J-gates. Also, a closer look into FIG.~\ref{Fig2}(b) reveals that J23, being defined in the first layer, actually forms a dot underneath easily with a relatively small increase in voltage.  

The four measurements in FIG.~\ref{Fig2} were performed with similar SET settings. In these measurements, we can compare the charge sensing signals as a function of pairwise double dot orientation and distance from the SET (see Figure~\ref{Fig2} center inset). From deducing the charge transition visibility, we find that the charge sensor has the best sensitivity on the nearest P3$-$P4 dots and worst sensitivity on the farther P2$-$P3 dots. Moreover, the SET is most sensitive to inter-dot tunneling events between dots when they (P1-P2 and P3-P4) are positioned perpendicular to the sensor axis (direction of current flow), as compared to parallel (P2-P3 and P1-P4), due to the increased dipole moment when the electron moves toward/away from the SET.

%begin{figure*}%[ht!]
%centering
%includegraphics[width=6.1in]{Fig1.pdf}

\section{Inter-dot tunnel rate measurements}\label{tunnel rate}

To assess the J-gate controllability, we perform pairwise inter-dot tunnel rate measurements. In FIG.~\ref{Fig3}(a), we focus on the (4,1)--(5,0) electron transition of the P3$-$P4 double dot charge stability diagram. Figure~\ref{Fig3}(b) illustrates the conduction band profiles for P3$-$P4 dots at high and low J34 voltages. At low $V_\mathrm{J34}$ (purple star), J34 forms a high tunnel barrier and prevents the electron tunneling between the P3 and P4 dots. The charge transition is barely visible indicating that the tunnel rate is significantly smaller than the excitation frequency applied to the gate. As $V_\mathrm{J34}$ increases, the tunnel barrier decreases and the likelihood of an electron tunneling event occurring between the dots increases. At high $V_\mathrm{J34}$ (marked yellow square), the tunnel barrier is low and it allows the transfer of electrons between the dots. Thus, a clear charge transition line is observed.

To probe the inter-dot tunnel rates, we measure the SET differential current, $\Delta$$I_\mathrm{D}$ as a function of gate-pulsed lock-in excitation frequency, $f$, along the charge transition line, as indicated by the red dashed line in FIG.~\ref{Fig3}(a). Figure~\ref{Fig3}(c) shows an example of $\Delta$$I_\mathrm{D}$ vs excitation frequency, $f$, at $V_\mathrm{J34}$= 1.208 V after removing the ac coupling effects between the setup lines. Then, we fit $\Delta$$I_\mathrm{D}$ as an exponentially decaying form of $\Delta I_{D}(f)= I_{D} \text{exp}\left [ (-\frac{f}{r})^n \right ]$ where $r$ is the electron tunnel rates. We repeat the same measurement on J23, J34 and J41 by sweeping excitation frequencies and their voltages along the corresponding transition lines, $\Delta$$V_\mathrm{J}$, and fit the tunnel rates as plotted in FIG.~\ref{Fig3}(d). The tunnel rates depend exponentially on the J-gate voltages. J23, being the wider gate and fabricated in layer 1, has the largest controllability on the tunnel rates between P2$-$P3 dots with a rate of change of 30.2 $\pm$ 4.9 dec/V. In contrast, J41, being fabricated in the last layer and having a pointed profile at the bottom, has the weakest influence on the tunnel rates between P1$-$P4 dots, at a rate of 2.5 $\pm$ 0.2 dec/V. 

\change{This variability in the rate of change of the tunnel coupling is closely linked to the device fabrication process, particularly because the J-gates are deposited in different metallization layers. The simulation tools presented in the Supplementary emulate the EBL fabrication process of the 2$\times$2 array, allowing us to incorporate the impact of these complex features in our modelling (See supplementary FIG.~S1). Each additional metal layer is encapsulated by a thin film of aluminium oxide, which increases their separation from the silicon substrate where the dots are formed, thus systematically reducing their effectiveness in tuning the interdot potential barrier (see FIG.~S2). The screening of the potential by the gates deposited in previous metal layers can also contribute significantly to this decay. As a result, gates positioned in higher metal layers are expected to exhibit reduced tunability of tunnel couplings and exchange interactions. In the supplementary FIG.~S3 we present the results of AB-initio simulations for both parameters using with a path integral approach~\cite{cifuentes2023path,cifuentes2024bounds}. These results are in good qualitative agreement with the trends observed in Fig.~\ref{Fig3}\textbf{d}, further supporting the validity of this explanation. }

% The key feature inducing the variations in the tunnel rate control is that the J-gates are defined in different metalization layers \change{(See supporting simulations in supplementary figures FIG. S1-S3)}. Each additional metal layer is encapsulated by a thin film of aluminium oxide, which increases their separation from the silicon substrate where the dots are formed, thus systematically reducing their effectiveness \change{in tuning the interdot potential barrier (see FIG.~S3). The screening of the potential by the gates deposited in previous metal layers can also contribute significantly to this decay in the effectiveness of the exchange gates. The simulation tools presented in the Supplementary emulate the EBL fabrication process of the 2$\times$2 array allowing us to incorporate the impact of these complex features in our modelling (FIG.~S1).}

%\change{The main reason to explain the observed variability in the control of the tunel rates between the different gates is that these are deposited in different metalization layers. Gates at higher layers. Only . If this gap is two small or the oxide is two thick .  }
%We confirm this in simulations performed in a digital twin of a similar device, emulating the specific features of EBL litography.

\begin{figure*}[ht!]
\includegraphics[width=6.1in]{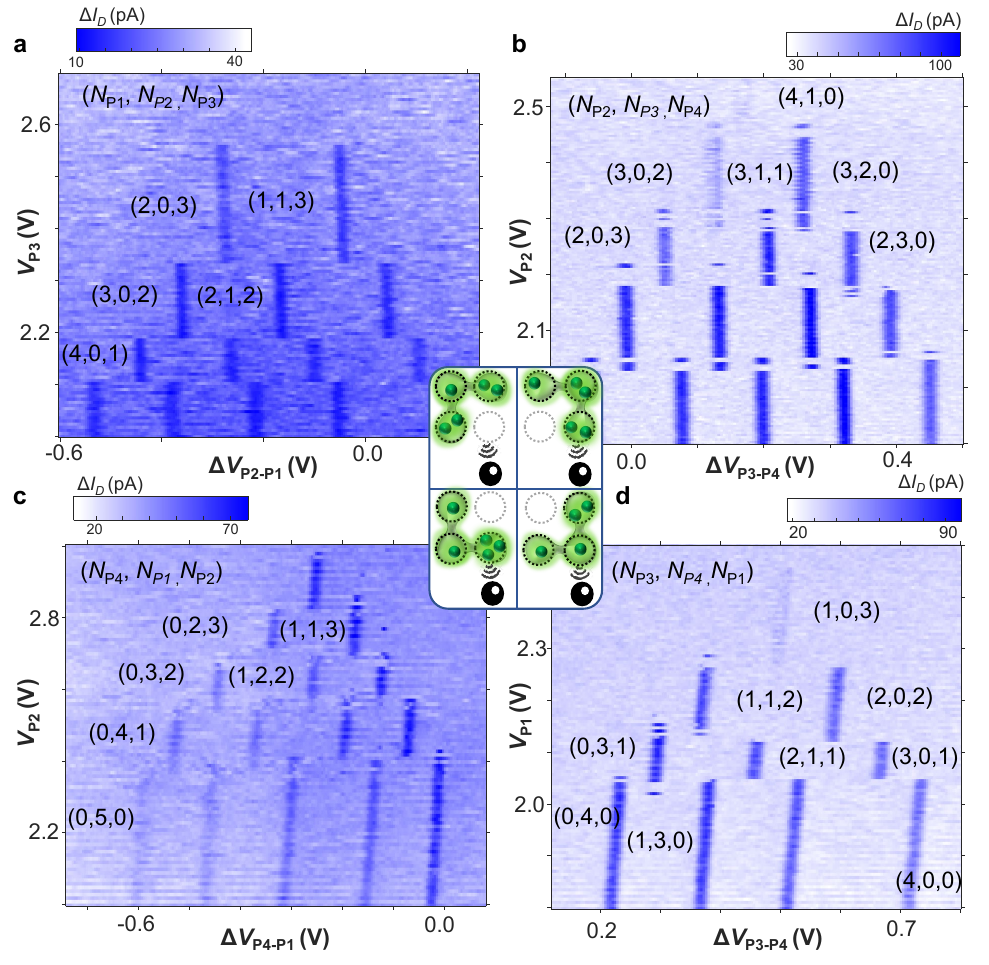}
\caption{Isolated-mode triple dot charge stability diagrams in four different configurations. Triple dot system of (a) P1$-$P2$-$P3, (b)P2$-$P3$-$P4, (c) P4$-$P1$-$P2, and (d) P3$-$P4$-$P1. Dots in operation are indicated in the inset. The electron occupancy on each dot is denoted as $N_\mathrm{P}$.   
}
\label{fig5}
\end{figure*}

\change{In this experiment, we were not able to assess spin parameters, such as the exchange interaction, because the measurements were performed at 4.2 K on a silicon substrate without nuclear spin purification—both of which are necessary for high-fidelity spin qubit demonstrations~\cite{tanttu2024assessment}. Nevertheless, we can still estimate the exchange control rate, as it is approximately twice the tunnel control rate $ \frac{d \log_{10}(J)}{dV_J} \approx 2 \frac{d \log_{10}(t)}{dV_J}$ (See details in the supplementary).  From the tunnel control rates $\frac{d \log_{10}(t)}{dV_J}$  measured in FIG.~\ref{Fig3}(d) of 2.5  dec/V for J41, 10.4 dec/V for J34, and 30.2 dec/V for J23, we can expect exchange rates of  5, 21 and 60 dec/V respectively. This allows us to predict if the device will meet the exchange controllability requirements without the need for dilution refrigeration. }
% The results obtained for the tunnel couplings can, however, allow us to get an approximation of the expected exchange control that this device would have, based on the formula:}

Typically an exchange control rate $\frac{d \log_{10}(J)}{dV_J}$ higher than $\sim$ 8 dec/V is desirable in order to be able to switch entanglement on and off despite the presence of random sources of variability (e.g. charge traps, Si/SiO$_2$ roughness, etc.)\cite{cifuentes2024bounds}. Under these assumptions, only J41 would fall below that threshold for a suitable J-gate design. \change{It is worth noting that the dot pitch in P1$-$P2 and P3$-$P4 is similar to the ones used in Ref. \cite{tanttu2024assessment}, where high exchange rates were achieved to perform two-qubit gates without accumulating J-dots.}

\change{A possible pathway to enhance the tunability of the J41  gate is increasing the pitch between the corresponding plunger gates (P1 and P4). See FIG.~S4 in the supplementary material.  The gate J41 deposited in the last layer would then fall deeper in the gap between P1 and P4 during the deposition, enabling better control of the interdot potential. }

% Add this package to the document preamble:
% For Beamer presentations, use \usepackage{colortbl} instead

\section{Triple Dot Charge Characteristics}\label{Triple Dot Charge Characteristics}

In addition to double dot configurations, we are also able to tune the device into triple dot configurations. Figure~\ref{fig5} shows the four charging maps of isolated mode triple dot systems in (a) P1$-$P2$-$P3, (b) P2$-$P3$-$P4, (c) P4$-$P1$-$P2, and (d) P3$-$P4$-$P1. The number of electrons occupying each dot is indicated by $N_\mathrm{P}$. In the P3$-$P4$-$P1 triple dot system (see FIG.~\ref{fig5}(d)), we first load four electrons into the P4 dot and form a (0,4,0) charge configuration at the bottom left of the plot. By increasing the detuning voltage $\Delta V_{P3-P4}$, we move electrons one by one into the P3 dot, from (0,4,0) to (1,3,0), through to (4,0,0). Similarly, by increasing the voltage on P1, we can move electrons one by one into the P1 dot, from (4,0,0) to (0,0,4). Note that the horizontal charge transitions are hardly visible because the tunnel rate between P1 and either of P3 or P4 is much lower than the gate-pulsed excitation frequency. To put it simply, we sweep the detuning voltage of two plunger gates and step the voltage of the third plunger gate, such that the third quantum dot acts as an electron reservoir from (to) where electrons can (un)load, forming the "Christmas tree" charging diagram. The inset in the centre of FIG.~\ref{fig5} illustrates the active triple dots in each measurement plot. Notably, the charge transition visibility is lowest for the P1$-$P2$-$P3 triple dot as it is furthest away from the SET. These measurements demonstrate the high tunability and stability of all the four dots in the 2$\times$2 quantum dot array.

\section{Conclusion and Outlook}
We have fabricated a 2D array of silicon MOS quantum dots and demonstrated pairwise double-dot and triple-dot arrangements in isolated mode at 4.2 K. With the four-layer gate stack, the device can be operated in the few-electron occupancy regime and possesses excellent tunability. The experiments offer important learnings on (i) the design of J-gate and its tunnel-rate controllability; (ii) the position of SET with respect to dots and its sensitivity and (iii) the characterisation of tunnel rates and its relation to exchange rates. \change{The simulation techniques presented in the Supplementary can help with future device design optimisation, ensuring high-level exchange controllability for qubit entanglement.}

\change{This demonstration underscores the feasibility of 2D MOS spin qubit devices. However, further scaling will require adopting technologies that enable the expansion of the array farther into the two dimensions. This could be achieved by implementing alternative readout methods, such as gate-based readout or single-electron-box sensing~\cite{west_gate-based_2019, SingleElecBoxQuantumMotion}, which occupy less space than current SETs. The in-plane antenna could also be replaced by an off-chip dielectric resonator~\cite{Vahapoglu2020}. Finally, SLEDGE architectures will enable the vertical fan-out of the gates forming the quantum dot array~\cite{ha2025twodimensionalsispinqubit, li2025trilinearquantumdotarchitecture}. In this context, the main contribution of the paper is outlining the necessary benchmarks for gate pitch, tunnel coupling controllability, and sensitivity to interdot transitions  to achieve a successful demonstration of a 2D spin qubit array in MOS technology.}

\vspace{.2in}\noindent{\textbf{SUPPORTING INFORMATION}}

\noindent
\change{FIG. S1: Simulation process to generate 3D model used for electrostatic simulations.
FIG. S2: Electrostatic simulations of the device.
FIG. S3: Path integral simulations of tunnel and exchange couplings.
FIG. S4: Exchange control rate vs gate pitch.}

\vspace{.2in}\noindent{\textbf{DATA AVAILABILITY}}

\noindent
The data supporting the findings in this study are available from the corresponding
authors upon reasonable request.

\vspace{.2in}\noindent{\textbf{ACKNOWLEDGEMENTS}}

\noindent
 We acknowledge support from the Australian Research Council (FL190100167, CE170100012 and IM230100396), the US Army Research Office (W911NF-17-1-0198, W911NF-23-10092), and the NSW Node of the Australian National Fabrication Facility. The views and conclusions contained in this document are those of the authors and should not be interpreted as representing the official policies, either expressed or implied, of the Army Research Office or the US Government. The US Government is authorized to reproduce and distribute reprints for Government purposes notwithstanding any copyright notation herein. This project was undertaken with the
assistance of resources and services from the National Computational Infrastructure (NCI), which is supported by the Australian
Government and includes computations using the computational
cluster Katana supported by Research Technology Services at
UNSW Sydney.

\vspace{.2in}\noindent{\textbf{Corresponding Authors}}

\noindent
Correspondence to the first or last authors.

\vspace{.2in}\noindent{\textbf{Competing Interests}}

\noindent
A.S.D. is the CEO and a director of Diraq Pty Ltd. W.H.L., T.T., A.D., K.W.C, F.E.H., C.C.E., C.H.Y., A.L., A.S. and A.S.D. declare equity interest in Diraq Pty Ltd.

\newpage
\onecolumngrid
\appendix
\section{Device Modeling}\label{Exchange Modelling}
\begin{figure}[ht!]
\includegraphics[width=6.2 in]{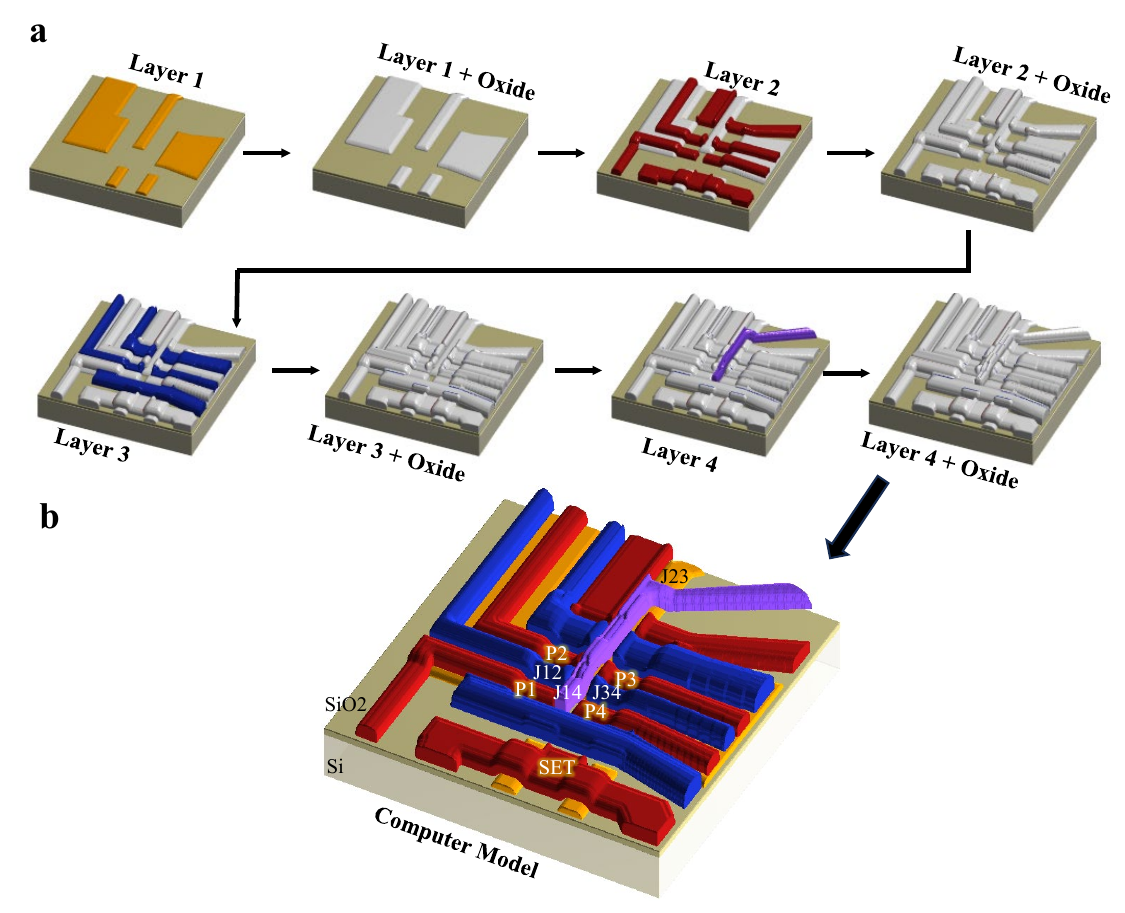}
\caption{\textbf{a,} Simulation of the device fabrication process described in FIG.~1 of the main text, emulating the thermal expansion of the four metal layers following deposition and subsequent oxide growth. \textbf{b,} Final 3D model imported into COMSOL Multiphysics for electrostatic simulations. The simulation parameters used to generate the 3D geometry are carefully tuned to closely match the physical device observed in  SEM images~\cite{cifuentes2024bounds} (see FIG.~1\textbf{a} of the main text).   }
\label{FigA1}
\end{figure}

\begin{figure}[ht!]
\includegraphics[width=6 in]{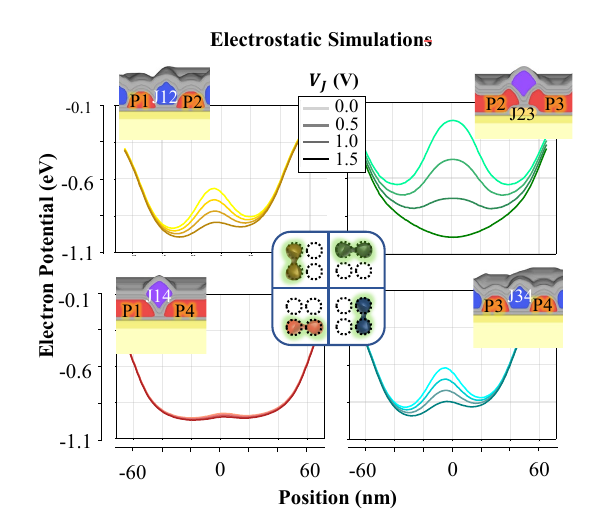}
\caption{ \textbf{a,} Cross section of finite element electrostatic simulations in COMSOL Multiphysics of the  potential of the computer model in FIG~\ref{FigA1}\textbf{b} for each double quantum dot pair (see inset schematics). The electric potentials are simulated with varying J-gate voltage showing distinct levels of effectiveness for each gate. }
\label{FigA2}
\end{figure}

\begin{figure}[ht!]
\includegraphics[width=5.5 in]{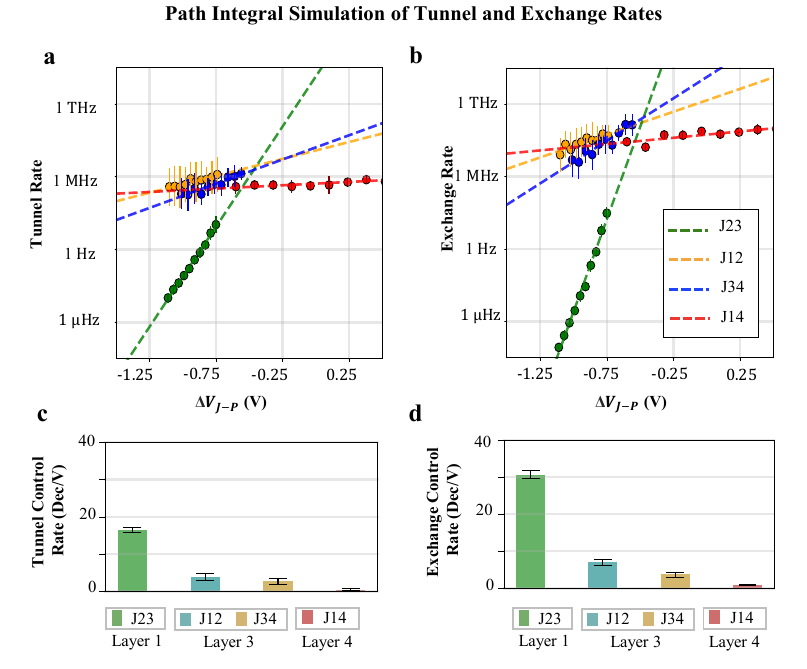}
\caption{ \textbf{a-b,} Path integral Monte Carlo simulations of \textbf{a)} tunnel couplings and \textbf{b)} exchange couplings as a function of J-gate biasing in the model of FIG.~\ref{FigA1}\textbf{b} (Example electrostatic potentials in FIG.~\ref{FigA2}).  $V_{J-P} = V_J - V_P$, where $V_P$ is the mean bias of the plunger gates defining the dots (See Ref. \cite{cifuentes2023path,cifuentes2024bounds} for details on the path integral simulation methods and bechmarking with experiments \cite{cifuentes2024bounds} and FCI simulations \cite{cifuentes2023path}). \textbf{c-d}, Electrostatic control rate of  \textbf{c)} the tunnel coupling $\left( \frac{d \log_{10}t}{dV_J} \right)$ and \textbf{d)} the exchange coupling $\left( \frac{d \log_{10}J}{dV_J} \right) $ by each J-gate in dec/V. The control rates are in qualitative agreement with the measurements in FIG.~3\textbf{d} of the main text with gate J23 being the most effective and  J14 the less effective. We note, however, that the values obtained for the  simulated tunnel control rates are significantly lower than experimental results limiting any quantitative comparison between experiments and simulations. This could occur because of the difficulty on reproducing  the device geometry  accurately (gate shape, granularity, variations in the oxide thickness, etc.) in the absence of TEM  images of the transversal cuts of the device as in Ref.~\cite{cifuentes2024bounds}. These features are critical to obtain a good estimate of the exchange control. }
\label{FigA3}
\end{figure}

The device modeling and simulation consists of three stages:
\begin{enumerate} \item Simulation of the device fabrication process (FIG.\ref{FigA1})
\item Finite-element simulation of the electrostatic potential in COMSOL Multiphysics (FIG.\ref{FigA2})
\item Ab initio path integral simulations of tunnel couplings and exchange interactions~\cite{cifuentes2023path} (FIG.~\ref{FigA3})
\end{enumerate}

\begin{figure}[ht!]
\includegraphics[width=3.2 in]{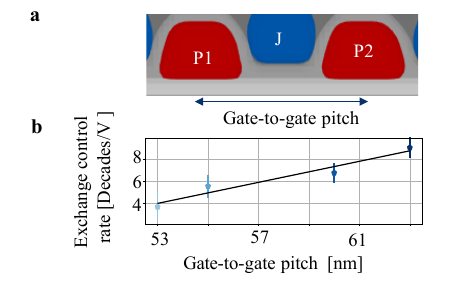}
\caption{ Simulation of the dependence of the exchange control rate on the inter-dot pitch. This simulation uses a simplified two-dot device model, distinct from the four-dot device shown in FIG.~\ref{FigA1}\textbf{b}. Although the device architecture differs, the fabrication process is similar, so the observed trend is expected to hold for the four-dot device as well.  }
\label{FigA4}
\end{figure}

Using this protocol, it is possible to predict the potential impact of changes in the device parameters, such as the gate-to-gate pitch (FIG.~\ref{FigA4}).

\section{Exchange Control Rate vs Tunnel Control Rate }
The exchange control rate can be approximated from the tunnel control rate as 
\begin{equation}
    \frac{d \log_{10}(J)}{dV_J} \approx 2 \frac{d \log_{10}(t)}{dV_J} 
    \label{eq:tunexch}
\end{equation}
This equation can be initially understood as the derivative of $J\approx \frac{4t^2}{U}$ with respect to $V_J$, derived from the Hund-Mulliken model \cite{burkard2023semiconductor}. Although this model may break down under strong interaction regimes, our previous experimental work \cite{tanttu2024assessment} - where tunnel and exchange rates were measured in the same device - suggests that Equation \eqref{eq:tunexch} remains valid in practice. In addition, in FIG.~\ref{FigA3}\textbf{c-d}, we present ab-initio path integral simulations of tunnel and exchange couplings for the four-dot system~\cite{cifuentes2023path,cifuentes2024bounds}, which also agree  with equation \eqref{eq:tunexch}. This method has been benchmarked against Full CI simulations ~\cite{cifuentes2023path} and experimental data~\cite{cifuentes2024bounds}, and is expected to remain valid at the strongly interacting regime, making it more reliable than the Hund-Mulliken model. This suggests that measuring tunnel-rate controllability at 4 K can serve as an effective predictor of whether a device will meet the exchange controllability requirements across all qubit pairs - without the need for dilution refrigeration.

%\beginsupplement
%\renewcommand{\theequation}{A\arabic{equation}}
%\renewcommand{\thefigure}{A\arabic{figure}}
%\setcounter{figure}{0}

%\pagebreak
%\widetext
%\begin{center}
%\large\textbf{Supporting Information: A 2x2 quantum dot array in silicon with tunable pairwise interdot coupling}\par
%\end{center}

% \renewcommand{\thefigure}{S\arabic{figure}}
% \def\theequation{S\arabic{equation}}

%\setcounter{figure}{0} 

%%\vspace{.2in}\noindent{\textbf{CONTRIBUTIONS}}
%%\noindent{\textbf{CONTRIBUTIONS}}

%\newpage
%\include{supp_int}
%\onecolumngrid
%\appendix

%\beginsupplement
%\renewcommand{\theequation}{A\arabic{equation}}
%\renewcommand{\thefigure}{A\arabic{figure}}
%\setcounter{figure}{0}

%\pagebreak
%\widetext
%\begin{center}
%\large\textbf{Supporting Information: A 2x2 quantum dot array in silicon with tunable pairwise interdot coupling}\par
%\end{center}

% \renewcommand{\thefigure}{S\arabic{figure}}
% \def\theequation{S\arabic{equation}}

%\setcounter{figure}{0} 

\twocolumngrid
%\newpage
\vspace{.2in}\noindent{\textbf{References}}
%\printbibliography[maxnames=5]
\bibliography{biblio_222}

\end{document}